\documentclass[useAMS,usenatbib,usegraphicx]{mn2e}
\usepackage{amssymb}
\usepackage{aas_macros}
\usepackage{color}

\def\iraf{\textsc{iraf}}
\def\ccdproc{\textsc{ccdproc}}
\def\kpnoslit{\textsc{kpnoslit}}
\def\rvsao{\textsc{rvsao}}
\def\molly{\textsc{molly}}
\def\doppler{\textsc{doppler}}
\def\elc{\textsc{elc}}
\def\ut{\textsc{ut}}

\newcommand\ion[2]{#1$\;${\scshape{#2}}}

\title[Optical spectroscopy of Cygnus~X-2]{Optical spectroscopy and Doppler tomography of Cygnus~X-2}
\author[P. Elebert et al.]{P. Elebert,$^{1}$\thanks{E-mail: p.elebert@ucc.ie}
        P. J. Callanan,$^{1}$
        M. A. P. Torres$^{2}$ and
        M. R. Garcia$^{2}$\\
$^{1}$Department of Physics, University College Cork, Cork, Ireland\\
$^{2}$Harvard--Smithsonian Center for Astrophysics, 60, Garden St., Cambridge, MA 02138, U.S.A.}

\begin{document}


\pagerange{\pageref{firstpage}--\pageref{lastpage}} \pubyear{2009}

\maketitle

\label{firstpage}

\begin{abstract}
   \noindent
   We present phase resolved optical spectroscopy and Doppler tomography of V1341~Cygni, the
   optical counterpart to the neutron star low mass X-ray binary Cygnus~X-2. We derive a radial
   velocity curve for the secondary star, finding a projected radial velocity semi-amplitude of
   $K_2 = 79 \pm 3$~km~s$^{-1}$, leading to a mass function of $0.51 \pm 0.06$~M$_{\sun}$,
   $\sim$30\% lower than the previous estimate. We tentatively attribute the lower
   value of $K_2$ (compared to that obtained by other authors) to variations in the X-ray irradiation
   of the secondary star at different epochs of observations. The limited phase
   coverage and/or longer timebase of previous observations may also contribute to the difference in $K_2$.
   Our value for the mass function implies a primary mass of $1.5 \pm 0.3$~M$_{\sun}$, somewhat lower than previous
   dynamical estimates, but
   consistent with the value found by analysis of type-I X-ray bursts from this system. Our
   Doppler tomography of the broad \ion{He}{ii}~$\lambda$4686 line reveals that most of the
   emission from this line is produced on the irradiated face of the donor star, with little emission from the
   accretion disc. In contrast, the Doppler tomogram of the \ion{N}{iii}~$\lambda$4640.64 Bowen
   blend line shows bright emission from near the gas stream/accretion disc impact
   region, with fainter emission from the gas stream and secondary star. This is the first
   LMXB for which the Bowen blend is dominated by emission from the gas stream/accretion
   disc impact region, without comparable emission from the secondary star. This has implications
   for the interpretation of Bowen blend Doppler tomograms of other LMXBs for which the ephemeris
   may not be accurately known.
\end{abstract}

\begin{keywords}
   accretion, accretion discs --
   stars: individual: Cygnus~X-2, V1341~Cygni --
   stars: neutron --
   X-rays: binaries
\end{keywords}

\section{Introduction}
 
Characterising the mass spectrum of neutron stars (NS) is critical to
the understanding of the equation of state (EoS) which describes
degenerate nuclear matter \citep{lattimer2004,lattimer2007}. Low mass X-ray binaries (LMXBs)
with NS primaries offer a promising means for
increasing the number of such mass estimates, and better constraining the
NS EoS. For example, \citet{vandenheuvel1995} have shown that some neutron stars
could accrete up to 0.7~M$_{\sun}$ over the course of the binary lifetimes. The detection
of such massive neutron stars would significantly reduce the range of currently
allowable NS equations of state.

LMXBs themselves are sub-divided into transient and persistent
systems. Persistent systems are those where the 
primary is continuously accreting at a significant fraction of
the Eddington limit, whereas in transient systems,
occasional outbursts are separated by long periods of quiescence,
generally explained by the disc instability model
\citep*{dubus2001}.
 
In quiescent transient systems, the mass of the compact object can be
constrained by dynamical studies if the secondary is bright enough. Spectroscopic
observations allow both the
orbital period and the velocity of the secondary (projected onto the line of sight) to be
determined, and these quantities are all that are required to calculate
the mass function ($f(M)$), a lower limit to the mass of the primary star. If the
optical counterpart is sufficiently bright to allow higher resolution
spectroscopy to be performed, the mass ratio can be estimated by
measuring the rotational broadening of the absorption lines
originating in the secondary star \citep*[e.g.][]{marsh1994}. Finally, fitting
of light curves can place constraints on the system's inclination,
providing the final piece of information required to constrain the
primary mass.

However, in LMXBs where the optical light is dominated by the accretion disc, other techniques are
required to find the system parameters.
Specifically, \citet{steeghs2002} found that in the case of
Scorpius X-1, the Bowen blend emission (a blend of \ion{N}{iii} and
\ion{C}{iii} emission lines near $\lambda$4640) can be used to
trace the motion of the secondary star. Doppler tomography
of Bowen blend
emission lines revealed a bright spot of emission, attributed to Bowen emission
from the irradiated surface of the secondary star. This
technique has since been used to determine the secondary velocity in a
number of other systems \citep[][and references therein]{casares2004,cornelisse2008},
although we note that no independent verification of $K_2$ has been
obtained for any of these systems to date.

Cyg~X-2 is a persistent NS LMXB, discovered in 1965 \citep{bowyer1965}, and is
the second X-ray binary for which an optical counterpart (V1341~Cygni) was found \citep{giacconi1967}. 
\citet{orosz1999} find that Cyg~X-2 is located at a distance of $\sim$7~kpc
and the high X-ray flux implies that it is accreting at close to the
Eddington limit. It is one of the few persistent systems where the secondary is visible, contributing
70\% of the optical flux \citep{orosz1999}.
Although the spectral type is A9~III \citep*{casares1998},
the mass of the secondary ($M_2$) is $\sim$0.6~M$_{\sun}$ \citep[see][for more details]{podsiadlowski2000}.
From the radial velocity (RV) curve based on absorption line spectra from the stellar
photosphere, \citet{casares1998}
established that $P_{\mathrm{orb}} = 9.8444 \pm 0.0003$~d and 
$K_2 = 88.0 \pm 1.4$~km~s$^{-1}$, yielding a value for the mass function of the NS in this system of
$f(M) = 0.69 \pm 0.03$~M$_{\sun}$. They also measured the systemic velocity ($\gamma$) to be $-209.6 \pm 0.8$~km~s$^{-1}$.
Their estimate for the rotational broadening of the secondary star of
$34.2 \pm 2.5$~km~s$^{-1}$ constrains
the mass ratio, $q$, to be $0.34 \pm 0.04$. Setting the lower limit of $M_2$ in line with the predictions
of \citet{king1997} allowed
\citet{casares1998} to constrain the primary mass ($M_1$) to be greater than 1.88~M$_{\sun}$, and
the orbital inclination ($i$)
to be less than 61$^{\circ}$, with 95\% confidence.
The existence of such a massive NS suggests that a stiffer EoS describes degenerate nuclear matter.

Subsequently, by fitting $U$-, $V$- and $B$-band light curves, \citet{orosz1999} found a value for
the inclination of $62^{\circ}.5 \pm 4^{\circ}$, which combined with the parameters found by \citet{casares1998} gave
a value for the primary mass of $1.78 \pm 0.23$ M$_{\sun}$. This latter mass estimate is consistent with the
canonical NS mass of 1.4~M$_{\sun}$ at the 2$\sigma$ level.
\citet{orosz1999} also found that X-ray heating of the secondary star contributed very little to the overall light
curves, possibly because of the large orbital separation, or because a flared accretion disc is shielding the secondary
from the X-rays.

In contrast, analysis of X-ray bursts from Cyg~X-2 allowed \citet{titarchuk2002} to constrain the primary mass
to be $1.44 \pm 0.06$~M$_{\sun}$, lower than that obtained by previous dynamical studies,
and equal to the canonical NS mass of 1.4~M$_{\sun}$, within the uncertainties.

Here, we present results based on spectroscopic observations of Cyg~X-2 taken
in 2006 September/October.  Our primary motivation for this
work was to perform Doppler tomography for this long period system, to
study the accretion flow, and especially to make the first comparison
between the radial velocity of the secondary as inferred from the
Bowen blend technique and absorption line measurements
\citep[i.e.][]{casares1998}. Confirmation of this technique for a
system where an independent measure of the radial velocity is
available gives further support for using Bowen blend to measure $K_2$
in situations where the secondary star is not directly observable.
In light of the NS mass estimates discussed above, we also
used our data to perform a new radial velocity analysis of this system
to confirm the suggested massive nature of the NS star primary.


\section{Data}
\label{data}

Our data consist of optical spectra, acquired with the FAST spectrograph \citep{fabricant1994},
at the Cassegrain focus of the 1.5-m Tillinghast telescope at the Fred L. Whipple Observatory, Mt. Hopkins,
Arizona. The FAST3 University of Arizona STA520A (SN4377) CCD was used, with $2720 \times 161$ 15~$\umu$m pixels
in the binned images.
The spectra were taken using the 600 lines~mm$^{-1}$ grating, with a 2~arcsec slit,
giving a spectral coverage of $\sim$2000~\AA.
The spectra were obtained on 16 nights in 2006 September and October, covering four orbits.
One spectrum was obtained each night, along with HeNeAr arc lamp exposures, bias, flat and dark frames, 
and one spectrophotometric standard exposure (BD$+28^{\circ} 4211$).
For 14 of the nights, only one arc spectrum was
available for the wavelength calibration of each Cyg~X-2 spectrum. For the other two nights, the Cyg~X-2
exposures were bracketed by arc spectra.

The frames were firstly processed
using the \iraf\footnote{\iraf\ is distributed by the National Optical
Astronomy Observatories, which are operated by the Association of Universities for Research in Astronomy, Inc.,
under cooperative agreement with the National Science Foundation.} \ccdproc\ routines to remove instrumental effects.
The spectra were then optimally extracted using
tasks in the \iraf\ \kpnoslit\ package. Wavelength solutions were found by fitting a 3rd order cubic spline to
$\sim$60 -- 70 lines in each arc spectrum, giving an RMS error of $<$0.04~\AA, and these solutions were then applied to the
target frames.
The resulting dispersion was $\sim$0.74~\AA~pixel$^{-1}$, with a resolution (measured from arc lines) of
$\sim$2.3~\AA. For comparison, the observations of \citet{casares1998} had a resolution of
$\sim$0.5 -- 0.8~A. Table \ref{observations} gives details of our observations.
In order to
verify the wavelength calibration, the sky lines were examined. Only five sky lines were
isolated and strong enough to be useful i.e. with a FWHM similar to the spectral resolution.

\begin{table}
\begin{center}
\centering
\begin{minipage}{0.45\textwidth}
\caption{Observations of Cyg~X-2 from 2006 September and October.}
   \label{observations}
   \begin{tabular}{| l | l@{}r |  l |  l | l |}
   \hline
   Date   & \multicolumn{2}{l}{Exp. time} & Coverage   & Dispersion  & Orbital \\
   (\ut)  & \multicolumn{2}{l}{(s)}  & (\AA)      & (\AA)       & phase $^{b}$ \\
   \hline
   Sep 17 $^{a}$ & & 1200 & 3456 -- 5453 & 0.743       &  0.98\\
   Sep 18 $^{a}$ & & 1200 & 3455 -- 5453 & 0.743       &  0.08\\
   Sep 19 & & 1200 & 3773 -- 5773 & 0.744       &  0.18\\
   Sep 20 & & 1200 & 3780 -- 5779 & 0.744       &  0.28\\
   Sep 21 & &  869 & 3779 -- 5780 & 0.744       &  0.38\\
   Sep 22 & & 1200 & 3773 -- 5773 & 0.744       &  0.49\\
   Sep 25 & & 1200 & 3770 -- 5770 & 0.744       &  0.78\\
   Sep 26 & & 1178 & 3774 -- 5774 & 0.744       &  0.89\\
   Oct 03 & & 1200 & 3775 -- 5775 & 0.744       &  0.60\\
   Oct 16 & & 1200 & 3787 -- 5786 & 0.744       &  0.92\\
   Oct 17 & & 1200 & 3786 -- 5785 & 0.744       &  0.02\\
   Oct 19 & & 1200 & 3785 -- 5784 & 0.744       &  0.22\\
   Oct 20 & & 1200 & 3788 -- 5788 & 0.744       &  0.33\\
   Oct 21 & & 1200 & 3768 -- 5768 & 0.744       &  0.42\\
   Oct 22 & & 1200 & 3780 -- 5780 & 0.744       &  0.52\\
   Oct 23 & & 1200 & 3784 -- 5784 & 0.744       &  0.62\\
   \hline
   \end{tabular}
\footnotetext{$^a$ Data on September 17 and 18 were taken with a slightly different central wavelength to the other nights.}
\footnotetext{$^b$ Based on the $T_0$ found by \citet{casares1998} and $P_{\mathrm{orb}} = 9.84456$~d.}
\end{minipage}
\end{center}
\end{table}

For nights where only a single arc spectrum was available, we found that there were significant
discrepancies between the measured wavelengths of these sky lines and
their rest wavelengths, as much as 0.5~ \AA. For those nights where the target observations
were bracketed by arcs, the interpolated wavelength solution resulted in
sky lines at their expected wavelengths.

\begin{figure*}
  \begin{center}
    \includegraphics[scale=0.44, angle=-90]{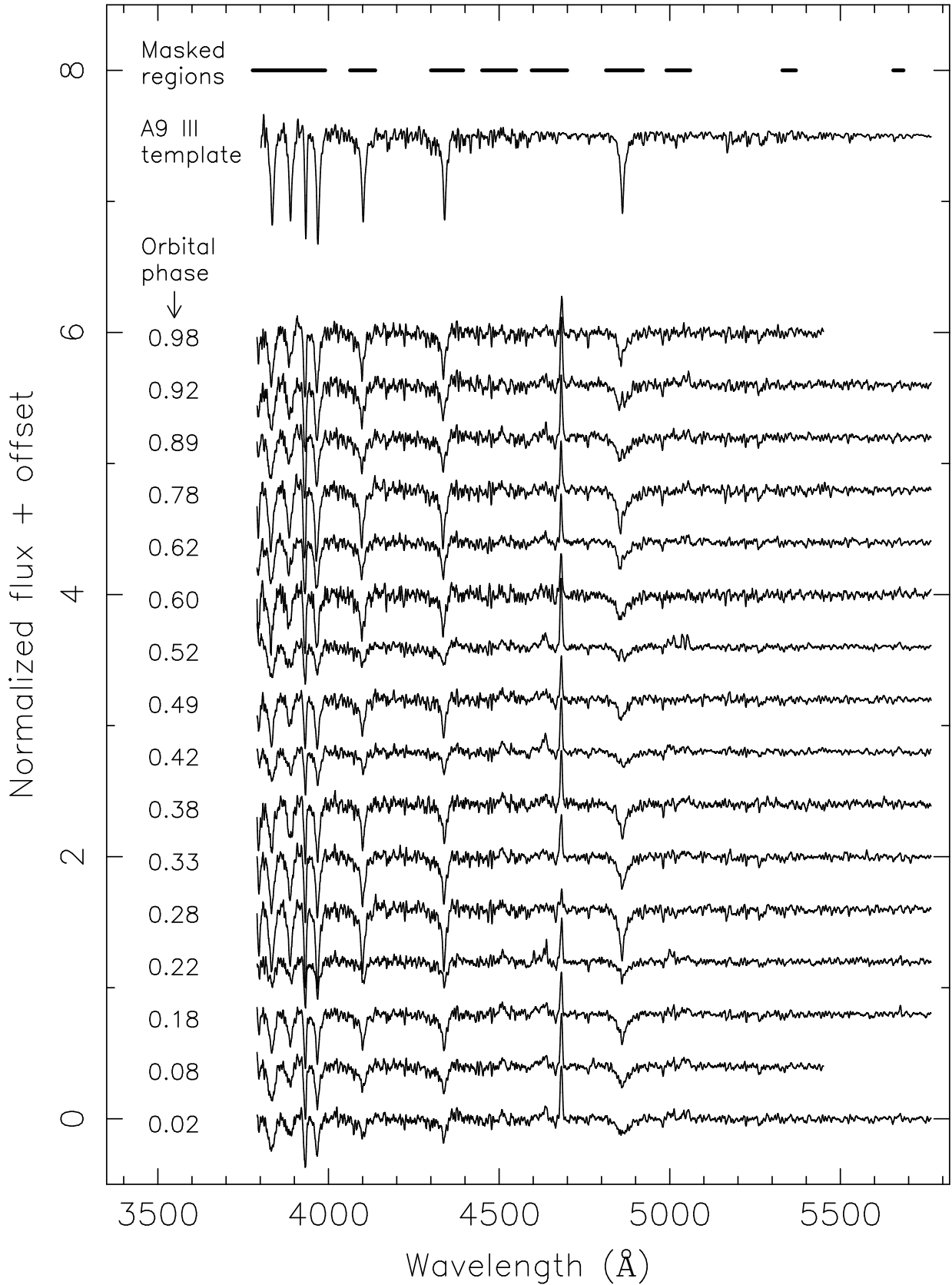}
    \caption{16 normalized spectra from 2006 September and October, with a vertical offset of 0.4 units.
	     The A9 III template spectrum is also shown. The
             wavelength regions masked from the cross-correlation are also marked. The spectra have been boxcar
             smoothed.}
    \label{norm}
  \end{center}
\end{figure*}

To correct for these offsets, we used the \rvsao\ package in \iraf\ \citep{kurtz1998} to cross-correlate
the sky spectra against the sky spectrum from October 16 (which was calibrated using two arc spectra)
and applied the measured wavelength shifts to the target spectra.
For the two nights where we had two arc spectra, we found that this technique
produced a wavelength scale very similar (maximum difference of $\sim$0.02~\AA) to that produced using the
two arc spectra.

After applying the sky line correction, the Cyg~X-2 spectra were imported into the
\molly\ spectral
analysis package, where the
spectra were shifted to the heliocentric frame, outlying flux values were cleaned and the spectra
were re-binned onto a common velocity
scale of 50~km~s$^{-1}$~pixel$^{-1}$. We found that the 1$\sigma$ errors produced by \iraf\ were significantly
lower than the RMS of continuum regions, and that this was different for each of the spectra. Therefore, we
increased the errors in each of the spectra to more closely match the RMS.
The spectra were rectified by fitting the continuum with a spline function
(after masking the major emission and absorption features), dividing by this fit and subtracting unity.
Fig. \ref{norm} shows these spectra, each with a vertical offset of 0.4 units.

We subsequently obtained spectra of 4 template stars in 2008 September,
spanning the range A5 to F1 III \citep[see][]{casares1998}.
The A9 III template (HR2489) is the same as that
used by \citet{casares1998}.
These frames were processed in an identical manner to the Cyg~X-2 frames. However,
arc spectra were taken at regular intervals, so the sky line correction procedure as employed for the
Cyg~X-2 spectra was not required. Inspection of the \ion{O}{iii}~$\lambda$5577.34 sky line showed that
it was located at the correct wavelength in each calibrated frame. Details of the template observations
are given in Table \ref{templates}. The normalized spectrum for the A9 III template HR2489 is also shown in
Fig. \ref{norm}.

\begin{table}
\begin{center}
\centering
\caption{Observations of template stars from 2008 September.}
   \label{templates}
   \begin{tabular}{| l | l |  l |  l | r |}
   \hline
   Date   & Template & Spectral & No.  & Total exposure  \\
   (\ut)  & star     & type     & obs. & time (s)        \\
   \hline
   Sep 24 & HD218260 & A5 III   & 3 & 180       \\
   Sep 24 & HD220999 & A7 III   & 4 & 60     \\
   Sep 24 & HD240431 & F1 III   & 3 & 270     \\
   Sep 25 & HR2489   & A9 III   & 9 & 165       \\
   \hline
   \end{tabular}
\end{center}
\end{table}

The spectra for each template were averaged and imported into \molly, where each spectrum was shifted
to the heliocentric reference frame, and the continua removed as for the Cyg~X-2 spectra.


\section{Results}

\subsection{Flux calibrated spectrum}

\begin{figure}
  \begin{center}
    \includegraphics[scale=0.36, angle=-90]{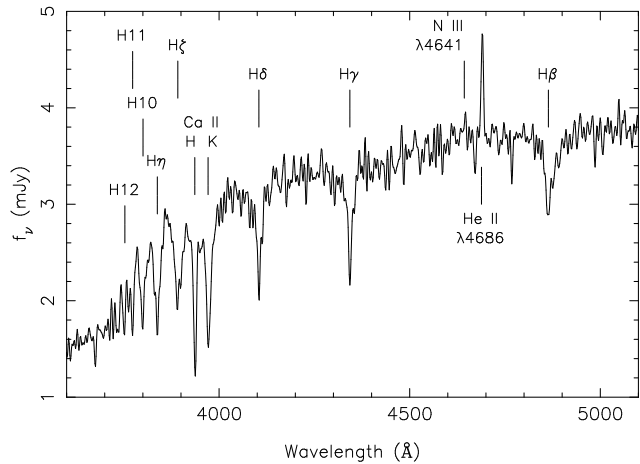}
    \caption{Flux calibrated spectrum from 2006 September 17 \ut. The spectrum has been shifted to the rest
             frame of the system, and the major spectral features have been marked.}
    \label{fluxcal1}
  \end{center}
\end{figure}

For spectra where observing conditions were photometric, flux calibration was performed against the standard star
BD$+28^{\circ} 4211$. Fig. \ref{fluxcal1} shows the flux calibrated spectrum from 2006 September 17 from which
we extracted a $B$-band magnitude of 15.2 mag. This is
consistent with the mean value over two decades reported by \citet{orosz1999} of 15.2 $\pm$ 0.2 mag (where the
error reflects the RMS error of all the values). Also, visual comparison of our spectrum with those of
\citet{vanparadijs1990} and \citet{obrien2004} shows that the spectrum is very similar at these epochs.

The equivalent widths (EW) and full-width half-maxima (FWHM) of the absorption and emission
lines in the average of all 16 spectra are shown in Table \ref{ew_fwhm}.

\begin{table}
\begin{center}
\centering
\caption{EW and FWHM of the emission and absorption lines in the averaged Cyg~X-2 spectrum.}
   \label{ew_fwhm}
   \begin{tabular}{| l | l | l | }
   \hline
   Line   & EW &  FWHM  \\
          & (\AA)  & (\AA)        \\
   \hline
   \ion{He}{ii}~$\lambda$4686  & $-2.9 \pm 0.1$ & $6.4 \pm 0.2$       \\
   Bowen blend                 & $-0.8 \pm 0.1$ & .......             \\
   H$\beta$                    & $5.2 \pm 0.1$  & $33 \pm 2$      \\
   H$\gamma$                   & $4.4 \pm 0.1$  & $18 \pm 2$      \\
   H$\delta$                   & $4.5 \pm 0.1$  & $18 \pm 2$      \\
   H$\zeta$                    & $3.6 \pm 0.1$  & $18 \pm 2$      \\
   H$\eta$                     & $6.1 \pm 0.1$  & $19 \pm 2$      \\
   \hline
   \end{tabular}
\end{center}
\end{table}

The strongest feature in the Bowen blend in the average spectrum is at a wavelength of
$\lambda$4637.4. Assuming a systemic velocity of $-210$~km~s$^{-1}$ \citep{casares1998}, we identify this
emission as from the \ion{N}{iii}~$\lambda$4640.64 line. This is frequently observed to be the strongest line in
the Bowen blend \citep{casares2004}. There does not appear to be any significant emission from the other
lines normally present in the Bowen blend. However, because of the resolution of our spectra, emission from
these lines may be present but blended with that from the \ion{N}{iii}~$\lambda$4640.64 line.

\subsection{Radial velocity study}
\label{rv_study}

As the primary motivation for this work was to produce Doppler tomograms,
we initially did not observe any radial velocity templates.
However, by using the individual spectra as templates, we were able to derive
a radial velocity curve, which suggested that $K_2$ was significantly lower than the value obtained
by \citet{casares1998}.
In order to confirm this lower value of $K_2$,
we later obtained spectra of
4 template stars, including the A9 III template used by \cite{casares1998}.
and we cross-correlated the Cyg~X-2 spectra against these templates to obtain
radial velocities.

\begin{figure}
  \begin{center}
    \includegraphics[scale=0.37, angle=-90]{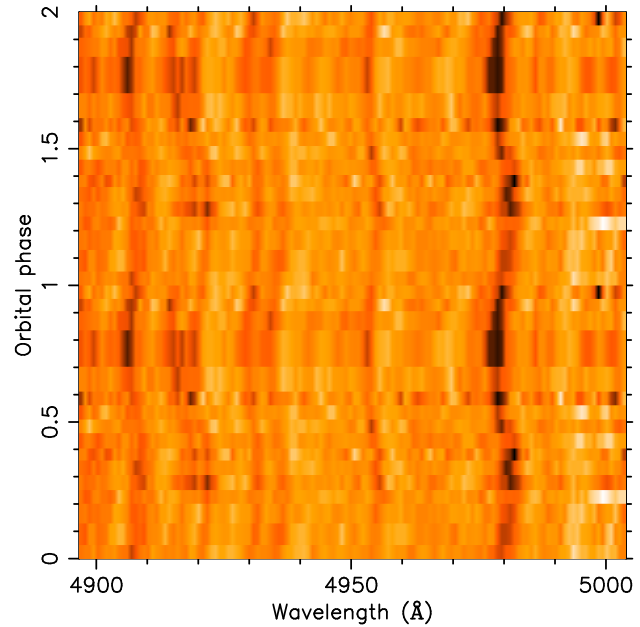}
    \caption{Trailed spectrogram for the data between $\lambda\lambda$4900 and 5000, showing
             some of the narrow absorption lines from the secondary.}
    \label{trailed}
  \end{center}
\end{figure}

For the cross-correlation, the broad Balmer absorption features, \ion{Ca}{ii} H and K lines, and
emission lines were masked (see Fig. \ref{norm}).
The trailed spectrogram in Fig. \ref{trailed} shows a subset of the data used in the cross-correlation
($\lambda$$\lambda$4900 -- 5000).
The S-waves due to the narrow absorption lines from the secondary are clearly visible.
The cross-correlations also took into account the uncertainties in the
wavelength solution of each
of the spectra, as well as the slight differences in the cross-correlation value when different
interpolations/binnings were used. This was accomplished by running 10000 cross-correlations for each spectrum-template
pair. For each cross-correlation, the wavelength scale of the spectrum and template were shifted independently by
choosing a shift value from a Gaussian distribution with a mean of 0 and standard deviation equal to the estimated
uncertainty in the wavelength solution due to the arc line fitting and the sky line correction.

\begin{figure}
  \begin{center}
    \includegraphics[scale=0.42, angle=-90]{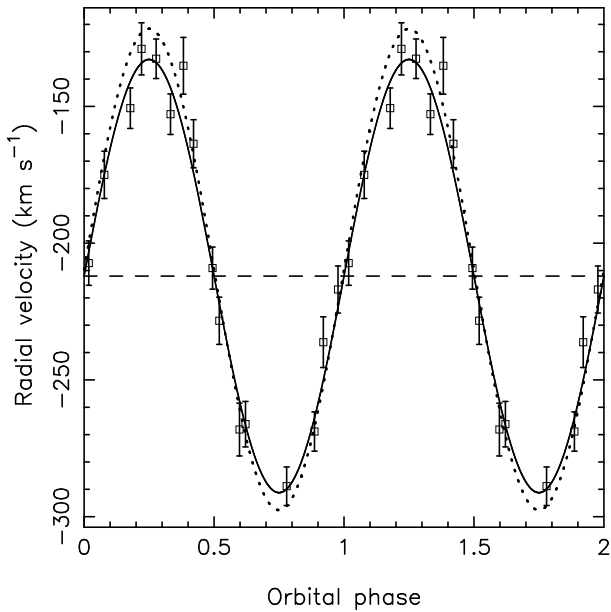}
    \caption{Radial velocity curve derived using the A9 III star HR2489 as a template. The best-fitting circular orbit
             is also plotted, and the $\gamma$ value of $-212$~km~s$^{-1}$ is marked with a dashed line. The curve
             of \citet{casares1998} is plotted with a dotted line. Two phases are shown for clarity.}
    \label{rv2}
  \end{center}
\end{figure}

We fit the 4 RV curves (derived from cross-correlation against the 4 templates) using
both circular and elliptical orbit models.
Using the \citet{lucy1971} test, we found that an elliptical orbit was not a
significant improvement over a circular orbit for any of these 4 RV curves: the eccentricity ($e$) in each case was
$\sim$$0.06 \pm 0.04$.
With a circular orbit and 3 free parameters ($K_2$, $\gamma$ and $P_{\mathrm{orb}}$, with $T_0$ set
to the value found by \citet{casares1998}), a sine fit to
each of these 4 RV curves gave
almost identical results (to better than 1$\sigma$).
Using the same A9 III template as used by \citet{casares1998}, we found
$K_2 = 79 \pm 3$~km~s$^{-1}$,
$P_{\mathrm{orb}} = 9.84456 \pm 0.00012$~d and
$\gamma = -212 \pm 2$~km~s$^{-1}$, where the radial velocity of the template of $7.7 \pm 0.1$~km~s$^{-1}$
\citep{gontcharov2006} has been included. This is in contrast to the values found by
\citet{casares1998}. For comparison, \citet{casares1998} found values of $K_2 = 88.0 \pm 1.4$~km~s$^{-1}$
 $P_{\mathrm{orb}} = 9.8444 \pm 0.0003$~d and $\gamma = -209.6 \pm 0.8$~km~s$^{-1}$.
The uncertainties in our measurements, here and in the remainder of the paper, are
1$\sigma$ and were determined using Monte Carlo simulations assuming Gaussian statistics.
The values we obtained for $P_{\mathrm{orb}}$ and $\gamma$ are consistent at approximately the 1$\sigma$
level with the values determined by \citet{casares1998}. However, the $K_2$ values differ at the 99.7\% level.
The RV curve obtained from cross-correlation
against the A9 III template star and best-fitting circular orbit are shown in Fig. \ref{rv2}.
The reduced $\chi^2$ ($\chi^2_\nu$) for this fit is 1.2 for 13 degrees of freedom. For comparison, the RV
curve generated using the parameters of \citet{casares1998} is also shown.

\subsection{System parameters}

Using the mass function equation and the values for $K_2$ and $P_{\mathrm{orb}}$ gives a value for the
mass function of $f(M)$ = $0.51 \pm 0.06$~M$_{\sun}$. Because of the higher resolution spectroscopy of
\citet{casares1998}, we use their value for the rotational broadening of the secondary star,
$v\sin{i}$ = $34.2 \pm 2.5$~km~s$^{-1}$. Combined with our lower value of $K_2$, we find a mass ratio
of $q$ = $0.42 \pm 0.06$. Along with our value for $f(M)$, we calculate $M_{1}\sin^{3}{i}$ = $1.03 \pm 0.15$~M$_{\sun}$.

X-ray eclipses have not been observed from this system, and this allows us to set a firm upper
limit to the inclination. Re-arranging the equation of \citet{paczynski1971}, we get:

\begin{displaymath}
	\frac{r_2}{a} = 0.462\left(\frac{q}{q+1}\right)^{\frac{1}{3}}
\end{displaymath}

\noindent
where $a$ is the binary separation and $r_2$ is the radius of the secondary star.

In order for the secondary not to eclipse the primary, the upper limit on the inclination
is $90^{\circ} - \tan^{-1}\frac{r_2}{a}$ which is 
$72^{\circ}.7 \pm0^{\circ}.6$. This angle is
relatively insensitive to changes in $q$. Using this limit, we place a lower limit on the primary mass
of $M_1$ $>$ $1.18 \pm 0.17$~M$_{\sun}$.

\citet{orosz1999} find a lower limit to the orbital inclination by fitting the ellipsoidal modulation in their $V$-band
light curve, assuming no contribution from the disc or from an irradiated secondary -- both of these would dilute the
observed modulation, and so the inclination found when these effects are not included is a lower limit. Taking the
published $V$-band data of \citet{orosz1999}, we used the \elc\ \citep[Eclipsing Light Curve:][]{orosz2000} code
to fit the modulation, using the parameters described in \citet{orosz1999}, and reproduced their value for the
inclination lower limit\footnote{Note that \citet{orosz1999} found lower limits of 42\degr\ and
49\degr\ for the $B$ and $V$-band light curves, respectively, and so adopted 49\degr\ as a lower limit.}
of $\sim$49\degr\ ($48^{\circ}.7 \pm 2^{\circ}.1$). 
We then adjusted the
\elc\ parameters for $P_{\mathrm{orb}}$, $f(M)$ and $q$
based on our spectroscopic results, and found the best fit was for $i = 49^{\circ}.8 \pm 2^{\circ}.0$.
This sets an upper primary mass limit of $2.3 \pm 0.4$~M$_{\sun}$.

We then re-fit both the $B$- and $V$-band light curves 
of \citet{orosz1999} with
\elc, with X-ray irradiation and the contribution from the accretion disc included.
We adjusted the parameters for $P_{\mathrm{orb}}$, $f(M)$ and
$q$ as before, and by optimizing the inclination and disc temperature
parameters\footnote{The equation in \elc\ which describes the disc
temperature as a function of radius is different to that used by
\citet{orosz1999}, so we therefore firstly adjusted the temperature
parameters to give the same disc temperature profile as used by
\citet{orosz1999} before optimizing these parameters.} found a
best-fitting inclination of $i = 63 \pm 3^{\circ}$, almost identical to the
value found by \citet{orosz1999} of $i = 62.5 \pm 4^{\circ}$. The uncertainty
was computed in a similar manner to \citet{orosz1999}. Combining this
inclination estimate with our value of $M_{1}\sin^{3}{i}$ = $1.03 \pm
0.15$~M$_{\sun}$, we obtained a primary mass of $M_1 = 1.5 \pm
0.3$~M$_{\sun}$, and a secondary mass of $M_2 = 0.63 \pm
0.16$~M$_{\sun}$.

\subsection{Doppler tomography}

Doppler tomography is an inversion technique where a velocity space image of the accretion disc, gas stream and
secondary star is generated using phase resolved spectroscopy \citep{marsh1988,marsh2001}; the
observed spectra are projections of the system at different orbital phases.
A Doppler tomogram is a two
dimensional image in velocity space ($V_{\mathrm{x}}$,$V_{\mathrm{y}}$), with intensity representing
strength of emission at a particular velocity.
The Doppler tomograms presented here were constructed using the maximum entropy method (MEM), as implemented in
the \doppler\ package.
Synthetic line profiles, created from an initial
tomogram of constant pixel values, were fitted to the observed line
profiles, by varying the pixel values. This process was continued
until the $\chi^2_\nu$ of the fit approached a target value.  As a
large number of tomograms can satisfy a particular target
$\chi^2_\nu$, the tomogram with the maximum entropy (i.e. the
smoothest image) was chosen. The entropy was defined relative to a
Gaussian blurred default image. We obtained similar tomograms using the
filtered back-projection technique implemented in \molly, where the
tomogram is created by smearing the filtered line profiles across the
image in directions determined by the orbital phases of the profiles.

\subsubsection{\ion{He}{ii}~$\lambda$4686 emission line}

\begin{figure}
  \begin{center}
    \includegraphics[scale=0.42, angle=-90]{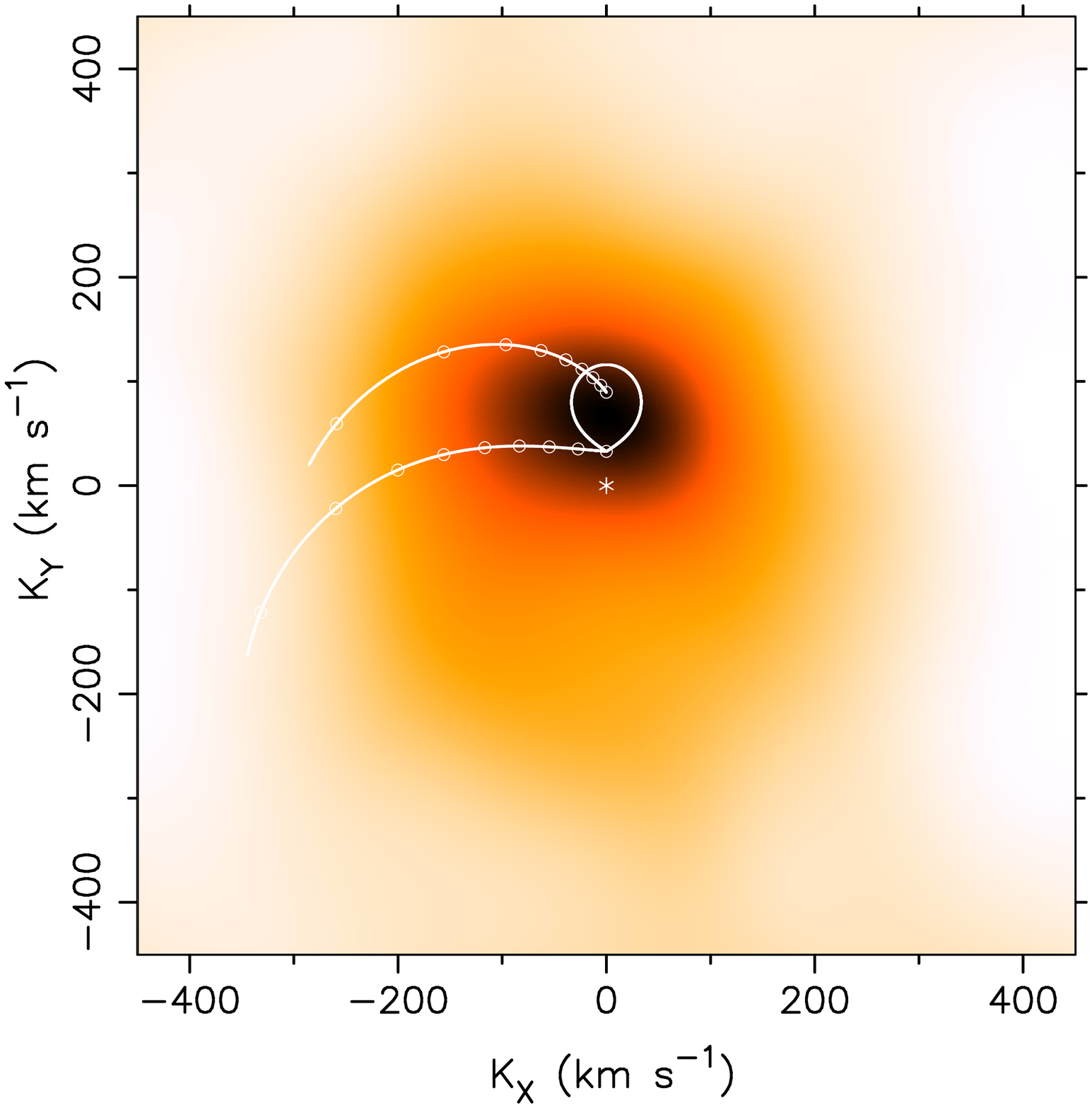}
    \caption{Doppler tomogram of the \ion{He}{ii}~$\lambda4686$ emission line. The gas stream and
             Roche lobe of the secondary are overplotted for $K_2 = 79$ km s$^{-1}$ and $q = 0.42$.}
    \label{dt_4686_full}
  \end{center}
\end{figure}

Fig. \ref{dt_4686_full} shows the Doppler tomogram for the \ion{He}{ii}~$\lambda$4686 emission line.
This emission line is often seen in X-ray binaries, and is formed by reprocessing of soft X-rays from
the central source.
The Roche lobe of the secondary is plotted, for $K_2 = 79$ km s$^{-1}$ and $q = 0.42$ (see Section \ref{rv_study}),
as is the centre of the mass
of the system, marked by a $\ast$. The lower curve is the gas stream velocity from the inner Lagrangian point,
and the upper curve is the velocity of the accretion disc along the gas stream.
The main feature to note is that most of the emission comes from near the secondary star, although
there is also a much broader
region of faint emission, due to the accretion disc.
with a distribution typical of that seen in accretion discs in other LMXBs.

\begin{figure}
  \begin{center}
    \includegraphics[scale=0.43, angle=-90]{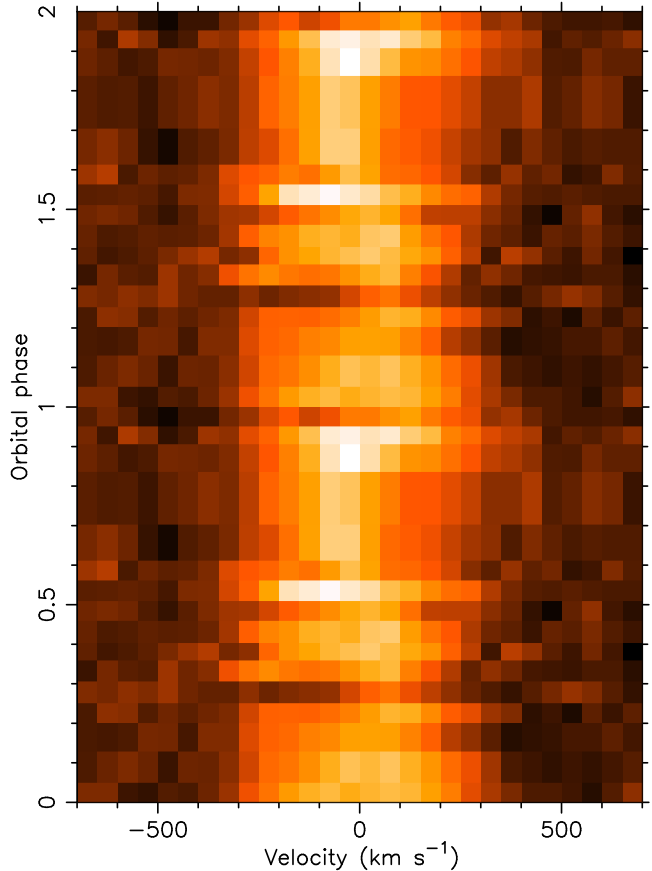}
    \caption{Trailed spectrogram of the data used to construct the \ion{He}{ii} Doppler tomograms.}
    \label{trail_4686}
  \end{center}
\end{figure}

\begin{figure}
  \begin{center}
    \includegraphics[scale=0.42, angle=-90]{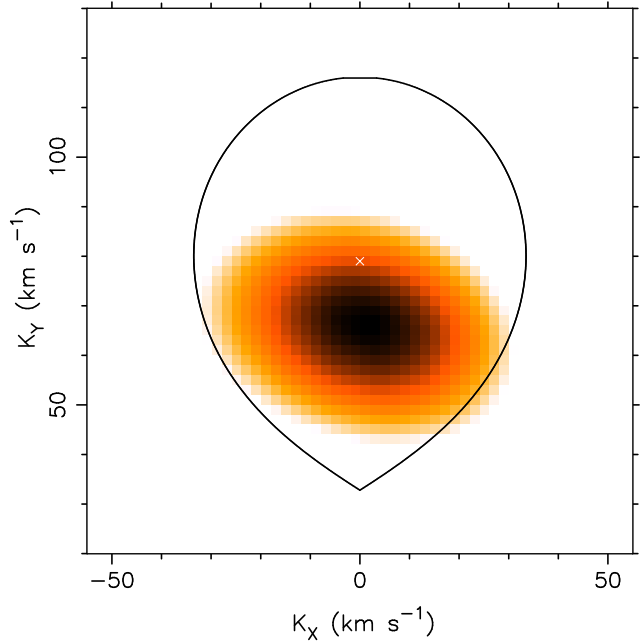}
    \caption{Doppler tomogram of the \ion{He}{ii}~$\lambda4686$ emission line, focussed on the region
             of the secondary star. The Roche lobe of the secondary is overplotted for $K_2 = 79$~km~s$^{-1}$
             and $q = 0.42$. The position of the secondary is marked with an error bar.}
    \label{dt_4686}
  \end{center}
\end{figure}

Fig. \ref{dt_4686} also shows the Doppler tomogram for the \ion{He}{ii}~$\lambda$4686 emission line,
focussing on the emission from near the secondary star -- the colour bar has been scaled to show only the peak emission.
This tomogram shows that the peak of the emission arises on the inner face of the secondary star.
The centroid of this emission is at a velocity of $K_{\mathrm{y}} = 64 \pm 15$~km~s$^{-1}$.

\subsubsection{\ion{N}{iii}~$\lambda$4640.64 emission line}

\begin{figure}
  \begin{center}
    \includegraphics[scale=0.37, angle=-90]{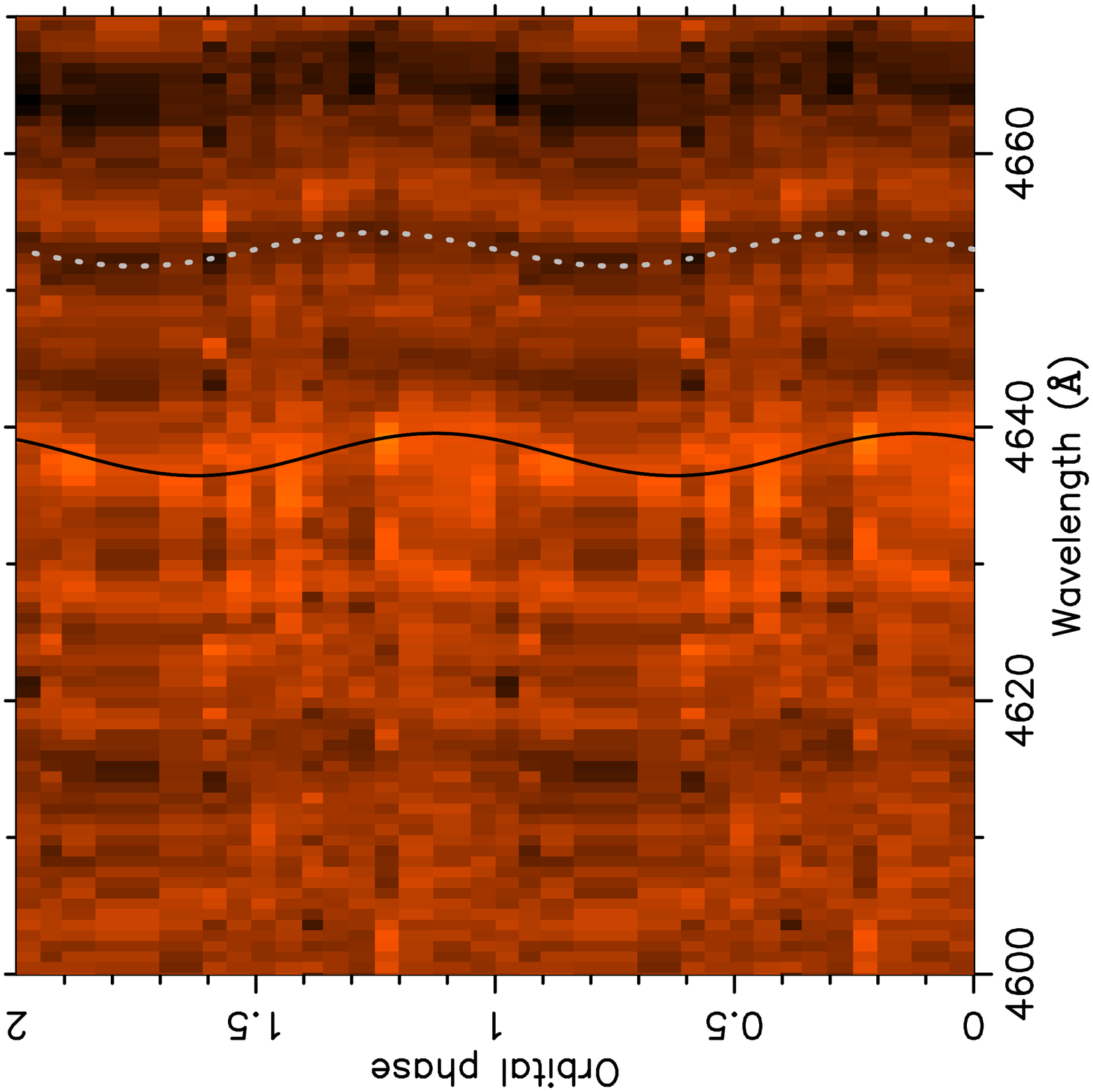}
    \caption{Trailed spectrogram of the data near the \ion{N}{iii}~$\lambda$4640.64 emission line. 
             The systemic velocity has not been removed, so the \ion{N}{iii}~$\lambda$4640.64 line is
             near $\lambda$4638. The solid line marks the S-wave from \ion{N}{iii}~$\lambda$4640.64
             emission, while the dotted line marks one of the nearby secondary absorption lines.}
    \label{trail_bowen_main}
  \end{center}
\end{figure}

Fig. \ref{trail_bowen_main} shows the
trailed spectrogram for the region near $\lambda$4640. Because the systemic velocity has not been
shifted out, the \ion{N}{iii}~$\lambda$4640.64 line appears near $\lambda$4638. Although
the S/N is low, it does appear that the S-wave for this line leads some of the other S-waves
present, (which are due to the motion of the secondary star).

\begin{figure}
  \begin{center}
    \includegraphics[scale=0.42, angle=-90]{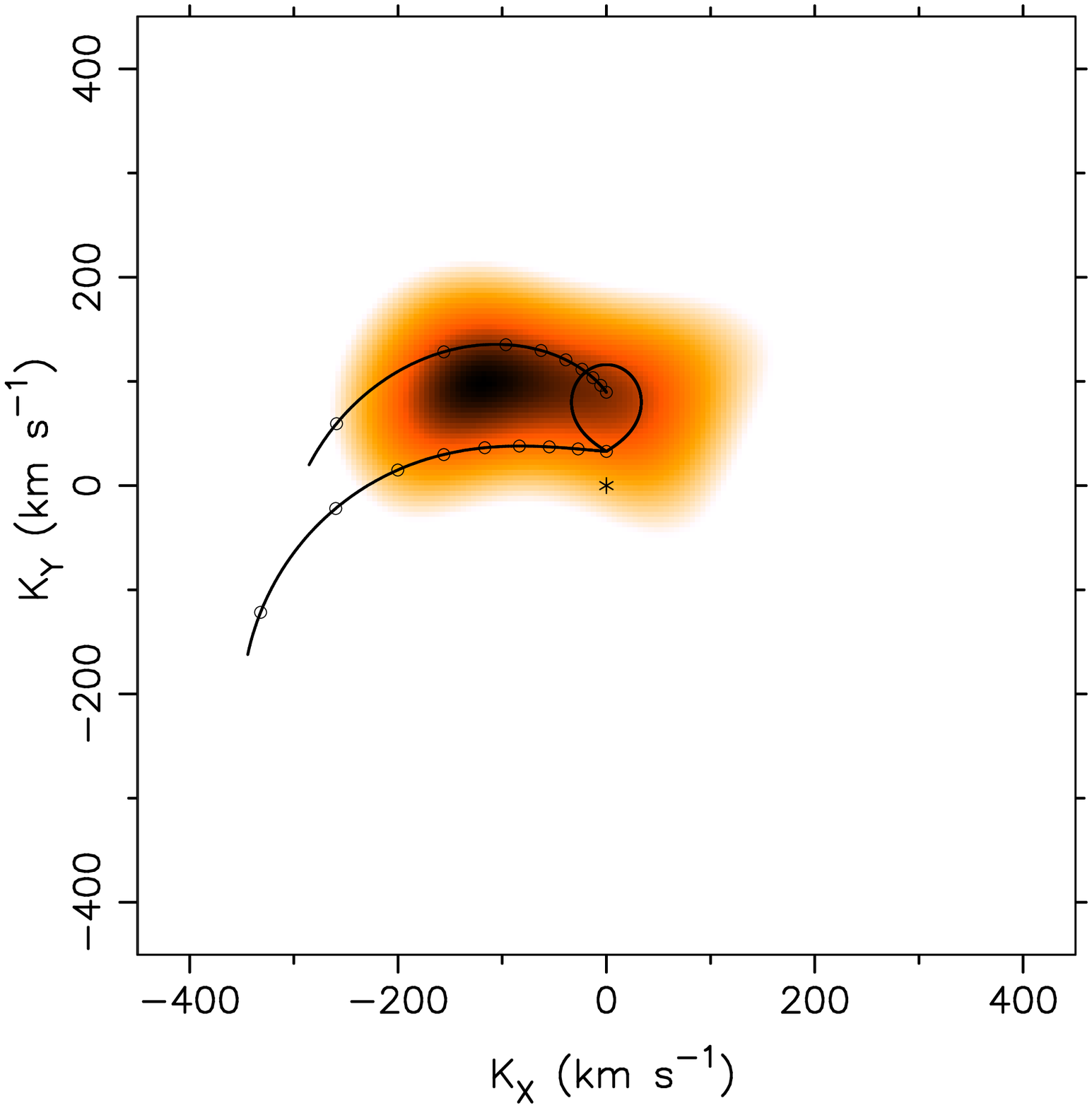}
    \caption{Doppler tomogram of the \ion{N}{iii}~$\lambda$4640.64 emission line. The gas stream
             and Roche lobe of the secondary are overplotted for $K_2 = 79$ km s$^{-1}$ and $q = 0.42$.}
    \label{dt_bowen_main}
  \end{center}
\end{figure}

Fig. \ref{dt_bowen_main} shows the corresponding Doppler tomogram, again with the Roche lobe and gas stream
overplotted. The strongest emission is from the location of the gas stream/accretion disc impact region,
with broad, fainter emission from the gas stream and near the secondary.
This is consistent with the situation we see in the
trailed spectrogram (Fig. \ref{trail_bowen_main}). The overplotted S-wave has a phase offset of
45$^{\circ}$, similar to what we see in the Doppler tomogram.

\section{Discussion}

\subsection{Projected velocity of the secondary star}

Our value for $K_2$ differs from the most recent observations of this system
by \citet{casares1998};
Table \ref{k2_gamma} summarises the values for $\gamma$ and $K_2$ found by various authors. Because
the work of \citet{casares1998} and our work show that the orbit is most likely circular,
we also re-fit the data published by \citet*{cowley1979} and \citet{crampton1980} assuming a circular
orbit.

\begin{table*}
\centering
\begin{minipage}{150mm}
\caption{$K_2$ and $\gamma$ found by different authors.}
   \label{k2_gamma}
   \begin{tabular}{| l | l | l | l | l | l | l |}
   \hline
   Source & Start of          & End of            & Time base & Method used & $K_2$          & $\gamma$           \\
          & observations      & observations      &           &             & (km s$^{-1}$)  & (km s$^{-1}$)      \\
   \hline
   a      & 1975 June 03      & 1978 August 09    & 3.2 yr    & Metal lines & $87 \pm 3$     & $-222 \pm 2$       \\
   b      & 1975 June 03      & 1978 August 09    & 3.2 yr    & Metal lines & $89 \pm 5$     & $-222 \pm 4$       \\
   c      & 1979 July 02      & 1979 September 17 & 77 d      & Metal lines & $80 \pm 6$     & $-203 \pm 3$       \\
   d      & 1979 July 02      & 1979 September 17 & 77 d      & Metal lines & $75 \pm 4$     & $-202 \pm 2$       \\
   e      & 1993 December 16  & 1997 August 07    & 3.6 yr    & Cross correlation & $88.0 \pm 1.6$ & $-209.6 \pm 0.8$   \\
   f      & 2006 September 17 & 2006 October 23   & 35 d      & Cross correlation & $79 \pm 3$ & $-212 \pm 2$       \\
   \hline
   \end{tabular}
\footnotetext{(a) KPNO data of \citet{cowley1979}.}
\footnotetext{(b) Based on our re-fitting of the KPNO data of \citet{cowley1979} assuming a circular orbit.}
\footnotetext{(c) \citet{crampton1980}.}
\footnotetext{(d) Based on our re-fitting of the data of \citet{crampton1980} assuming a circular orbit.}
\footnotetext{(e) \citet{casares1998}.}
\footnotetext{(f) This paper.}
\end{minipage}
\end{table*}

In our work, and in the work of \citet{casares1998} the non-metal lines and the \ion{Ca}{ii} lines were masked out
during cross-correlation (see Fig. \ref{norm}).
Therefore all values of $K_2$ and $\gamma$ in Table \ref{k2_gamma} are essentially based on the velocities of
the metal absorption lines.

We now examine possible reasons for our lower value of $K_2$ compared to that of \citet{casares1998}.
Firstly, it is possible that there are systematic errors in our wavelength
solution, given that for most of our spectra only a single arc lamp exposure was available, and a correction
based on the sky spectra was required. However, our RV curves
appear to be quite sinusoidal, which would be unlikely if each spectrum had random errors in the
wavelength calibration -- indeed, the RV curve from the uncorrected spectra deviates considerably
from a sinusoid.
Also as we mention in Section \ref{data}, for the cases where we had two arc spectra, the
maximum difference between the wavelength solutions using the two arc spectra, and using the sky line correction
technique is $\sim$0.02~\AA, or $\sim$1.3~km~s$^{-1}$ at $\lambda$4686.
Our RV curves include the uncertainties in wavelength scales, and also the
slight differences in the cross-correlation values that arise when different interpolations are used.
Finally, we find a value for $\gamma$ almost identical to that found by \citet{casares1998}, which
would again be improbable if there were any systematic errors in our wavelength scales.

Hence, we believe our measurement of $K_2$ to be accurate, and so we require an alternative explanation for the
difference between this measurement and that of \citet{casares1998}. 
Because our \ion{He}{ii} Doppler tomography confirms that the secondary is irradiated by X-rays from the central
source, the absorption lines from the irradiated face of the secondary star may be weaker than those from the
non-irradiated face. 
The irradiation of the secondary may be such that the absorption lines are completely quenched on
the irradiated hemisphere of the secondary star. In this case, an RV curve based on the velocity of absorption lines
from the secondary star will result in a $K_2$ value which reflects the velocity
of the non-irradiated hemisphere. This is larger than the velocity of the centre of mass of the secondary star.
If the irradiation of the secondary was variable, such that at some phases the absorption lines on the irradiated
face of the secondary were not fully quenched, then one would expect 
the $K_2$ value derived from the resulting RV curve to lie somewhere between the true value and the maximum value obtained
for complete quenching.
See \citet{wade1988} for further discussion of this effect.
\citet{steeghs2007} argue that differences in $K_2$ found by different authors for the LMXB 2S~0921$-$630 may be due to
variable irradiation effects, and we suggest that the same may well apply in the case of Cyg~X-2.
\citet{clarkson2003} show that the long term X-ray variability in Cyg~X-2 can be explained 
by a precessing warped accretion disc. The changing orientation of this disc warp will
cause the irradiation of the secondary to vary \citep[see also][]{vrtilek2003}.

To estimate the effects of irradiation for Cyg~X-2, we use the
approach of \citet{wade1988}. 
If the absorption centre of the secondary star is displaced from the centre of
mass by an amount $\Delta r = f r_2$, where $r_2$ is the radius of the secondary
star and $f$ is the fractional displacement, then:
\begin{displaymath}
   \Delta K_2 = \frac{\Delta r}{a_2} K_2
\end{displaymath}
where $a_2$ is the separation between the secondary and the binary centre of mass. Based on our new value
for $q$, $a_2 = 0.70a$. Using the formula of \cite{paczynski1971}, the
radius of the secondary is $0.31a$. Therefore, $\Delta K_2 = 0.443fK_2$.
In the extreme case, where there is no absorption
from the inner hemisphere of the secondary star,
$f = 4/(3\pi)$, and $\Delta K_2 = 0.19K_2 \simeq 15$~km~s$^{-1}$.
Therefore, we find that the measured value of $K_2$ can be as much as $\sim$15~km~s$^{-1}$ greater than the true value.
This difference is more than enough to explain the different values found by various authors.

\begin{figure}
  \begin{center}
    \includegraphics[scale=0.28, angle=-90]{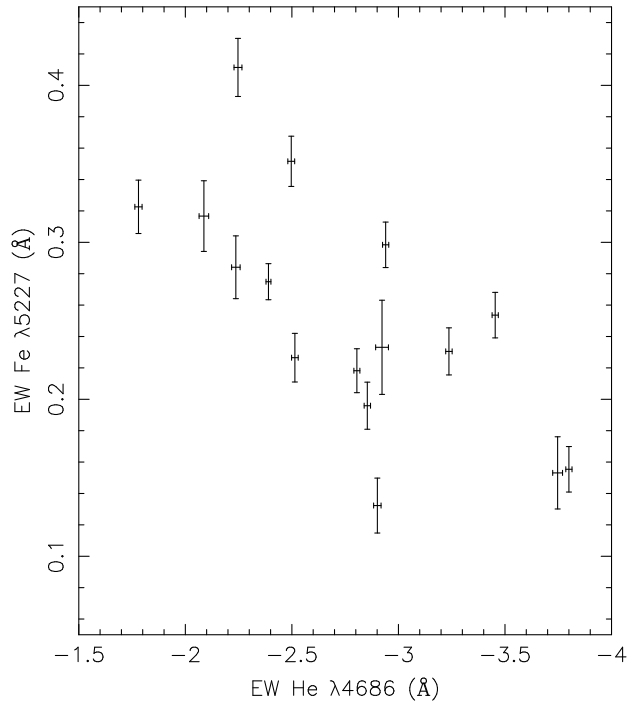}
    \caption{EW of the \ion{Fe}{i}~$\lambda$5227.15 absorption line vs. the EW of
             the \ion{He}{ii}~$\lambda$4686 emission line. Assuming that the
             EW of the \ion{He}{ii}~$\lambda$4686 emission line is correlated with the intensity of the X-rays
             intercepted by the secondary star, this relationship shows that the EW of
             the \ion{Fe}{i}~$\lambda$5227.15 absorption line decreases with increasing X-ray flux.}
    \label{ew}
  \end{center}
\end{figure}

In order to investigate the effect of irradiation on the absorption features of the secondary in Cyg~X-2, we show in
Fig. \ref{ew} the EW of the \ion{Fe}{i}~$\lambda$5227.15 absorption line as a function
of the EW of \ion{He}{ii}~$\lambda$4686 (which we take as indicative of the intensity of the X-rays intercepted by
the secondary star). This shows that, for this line at least, the stronger the
\ion{He}{ii}~$\lambda$4686 emission line (and hence X-ray flux), the weaker the secondary absorption line.

On the other hand, a Doppler tomogram of the \ion{Ca}{i}~$\lambda$4227 absorption line
shows that the peak occurs at a value consistent with that found in our RV analysis. However, the
EW of this line does not show any trend when plotted against the EW of the \ion{He}{ii}~$\lambda$4686 emission
line. This may suggest that this particular line is less susceptible to the effects of irradiation than
other lines (e.g. the \ion{Fe}{i}~$\lambda$5227.15 line), and that the radial velocity determined from this
line more accurately reflects the motion of the centre of mass of the secondary star.
Because irradiation of the secondary star causes the measured $K_2$ to be greater than the
true value, our measurement of a lower value may be closer to the true value than the estimates
of \citet{cowley1979} and \citet{casares1998}.

We also note from Table \ref{k2_gamma} that
observations obtained over a shorter time base \citep[][and this
work]{crampton1980} appear to suggest systematically lower values of
$K_2$ than those obtained over a longer period \citep{cowley1979,
casares1998}.  We speculate that observations obtained over a longer
timebase, and wider range of disc precession phase
\citep[e.g.][]{clarkson2003}, result in more variable irradiation of
the secondary star. It is possible that the changing values of $\Delta K_2$
that result from this may explain the variation in $K_2$ we observe.

\subsection{Primary mass and distance}

Our results indicate that the mass of the NS in Cyg~X-2 ($M_1 = 1.5 \pm 0.3$~M$_{\sun}$) is typical of a canonical NS, a
downward revision of the previous estimates of a 95\% confidence lower limit of 1.88~M$_{\sun}$ \citep{casares1998}
and $1.78 \pm 0.23$~M$_{\sun}$ \citep{orosz1999}.
It is interesting to note that the LMXB 2S~0921$-$630 has also recently had its primary mass estimate reduced,
from $3.2 \pm 1.2$ M$_{\sun}$ \citep{shahbaz2004} to $1.37 \pm 0.13$~M$_{\sun}$ \citep{shahbaz2007} and
$1.44 \pm 0.10$~M$_{\sun}$ \citep{steeghs2007}.
Our dynamical mass estimate for the NS in Cyg~X-2 is consistent with the mass estimate of \citet{titarchuk2002}
($M_1 = 1.44 \pm 0.06$~M$_{\sun}$), based on observations of type-I X-ray bursts from this system.

Calculations by \citet{orosz1999} found that there are inconsistencies between the distance based
on their optical observations ($d = 7.2 \pm 1.1$~kpc) and the distance estimate of \citet{smale1998} based on X-ray
observations of a type-I radius-expansion burst ($d = 11.6 \pm 0.3$~kpc). To address this, we
now re-calculate the distance and observed absolute visual magnitude ($M_{\mathrm{V}}$) of Cygnus X-2, and compare the
latter with the expected value based on our new parameters.

Taking the peak X-ray flux of $1.52 \times 10^{-8}$~erg~cm$^{-2}$~s$^{-1}$ \citep{smale1998} to be at the Eddington
limit, and with
$M_1 = 1.5 \pm 0.3$~M$_{\sun}$, the Eddington
luminosity of the system is $2.0 \pm 0.4 \times 10^{38}$~erg~s$^{-1}$ (assuming cosmic composition)
and the distance to the source is therefore $10.4 \pm 1.0$~kpc. With $V = 14.8$~mag, and $A_V = 1.24 \pm 0.22$~mag
\citep{mcclintock1984,orosz1999}, this gives $M_{\mathrm{V}} = -1.5 \pm 0.3$~mag for the system.

Assuming a secondary star with a surface temperature of $T_2 = 7000 \pm 250$~K \citep{orosz1999},
$M_1 = 1.5 \pm 0.3$~M$_{\sun}$ and $q = 0.42 \pm 0.06$, the radius of the Roche lobe of the secondary
star is $7.6 \pm 0.7$~R$_{\sun}$ \citep{eggleton1983}.
Hence, we calculate that the $M_{\mathrm{V}}$ of the secondary should be $-0.47 \pm 0.25$~mag. 
With a $V$-band disc fraction
($k_{\mathrm{V}} \equiv f_{\mathrm{disc}}/(f_2 + f_{\mathrm{disc}})$) of $0.30 \pm 0.05$ \citep{orosz1999}, the
absolute magnitude
of the system should therefore be $-0.86 \pm 0.26$~mag. This is marginally consistent with the
expected value for this system, calculated above assuming a distance of $10.4 \pm 1.0$~kpc. These estimates would better
agree if the secondary temperature was slightly higher. Indeed, we note that the
spectral classification of \citet{casares1998} is A9~$\pm$~2 subtypes, and an A7 star (with a surface temperature of
$7700 \pm 300$~K) would yield $M_{\mathrm{V}} = -0.88 \pm 0.26$~mag for the secondary and
$-1.27 \pm 0.27$~mag for the
system as a whole, leading to a distance of $9.4 \pm 1.5$~kpc. Hence, we conclude that the absolute magnitude calculated
from the observed apparent magnitude and the distance based on X-ray measurements is consistent with the expected
absolute magnitude based on the spectral type, surface temperature, disc fraction, primary mass and mass ratio.

For many systems, the secondary star dominates the infrared light, and fitting the ellipsoidal modulation in
the IR light curves can be used to better constrain the inclination. For Cyg~X-2, such a scenario would be
particularly desirable, as the modulation in the optical light curves is dominated by variable contamination
from the accretion disc. The \emph{2MASS} $K$-band magnitude of Cyg~X-2 is $\sim$13.05~mag. Using
$E(B-V) = 0.4$ \citep{mcclintock1984}, $A_{\mathrm{K}}/E(B-V) = 0.38$ \citep{savage1979} and a distance of
10.4~kpc, the dereddended absolute $K$-band magnitude is $M_{\mathrm{K}} = -2.2 \pm 0.2$~mag.
If the secondary is an A7 star, then $V-K \simeq 0.8$~mag. Using the absolute $V$-band magnitude for the secondary of
$-0.88 \pm 0.26$~mag (calculated above), $M_{\mathrm{K}}$ for the secondary should be $\sim$$-1.7 \pm 0.3$~mag. To
achieve the observed $M_{\mathrm{K}}$ for the system of $-2.2 \pm 0.2$~mag would require a $K$-band disc fraction of
$k_{\mathrm{K}} = 0.37^{+0.18}_{-0.25}$. Therefore, it would appear that even in the $K$-band, the accretion disc
contributes a considerable fraction of the light.

\subsection{Doppler tomography}

Our Doppler tomography shows that the peak of the \ion{He}{ii} emission originates on the inner hemisphere of the donor
star. This is consistent with the results of \citet{vrtilek2003}, who find that the emission
lines in their UV spectra also appear to originate in the secondary star.
There appears to be little or no emission from the gas stream/accretion disc impact region. 
Curiously, \citet{orosz1999} find that the optical light curves show no evidence for heating of the secondary star,
yet our \ion{He}{ii} Doppler tomograms show that the secondary is certainly intercepting X-rays from
the primary. It is unclear how the secondary could be irradiated by the central source, and yet the
light curves show no evidence for heating.

The situation is reversed in the \ion{N}{iii}~$\lambda4640.64$ tomogram (Fig. \ref{dt_bowen_main}) , where the
majority of the emission arises at the location of the hotspot, between the gas stream trajectory and the
velocity of the accretion disc along the gas stream.

Generally, the Bowen blend contains
several lines, with the strongest normally being the \ion{N}{iii}~$\lambda4640.64$ line.
However, in our data, this is the only line which is clearly visible.
The location of the emission
in the Doppler tomogram shows that significant Bowen blend emission can occur from sources other than
the secondary star in LMXBs. This is in contrast to the suggestion by \citet{cornelisse2008} that
the velocity of the Bowen blend emission can be used universally to determine the orbital
velocity of the secondary star.

\section{Conclusions}

By cross-correlating our spectra against template
spectra, we find a value for the projected radial velocity semi-amplitude of the secondary which is
significantly lower then that found by \citet{casares1998}.
Based on
this new value for $K_2$, we find a larger value for the mass ratio, using the rotational broadening value calculated by
\citet{casares1998}. Combined with the inclination estimates by \citet{orosz1999}, this implies that the mass of the
primary in this system is $1.5 \pm 0.3$~M$_{\sun}$, less massive than previously determined, although still
consistent with the mass estimate of \citet{orosz1999} at the 1$\sigma$ level, and similar to the value found
by \citet{titarchuk2002} based on observations of type-I X-ray bursts.

We have presented Doppler tomograms of Cyg~X-2, the first Doppler tomography of such a long period system. Our
tomography of the \ion{He}{ii}~$\lambda$4686 emission line shows that the majority of the emission is from
near the irradiated face of the secondary, with much weaker emission from 
the accretion disc. Doppler tomography of the \ion{N}{iii}~$\lambda$4640 emission line shows that the emission
is primarily from the gas stream/accretion disc impact region.
The \ion{N}{iii} tomography demonstrates that care must be taken when interpreting Bowen blend tomograms, so that the
hotspot is not confused with the secondary, particularly for systems where the ephemeris is not accurately known.

The measured value of $K_2$ can vary by up to 15~km~s$^{-1}$,
depending on the irradiation of the secondary star by the central
X-ray source, and this may explain the differences in $K_2$ found by
different authors. Further investigation of the
effects of irradiation of the secondary star -- for example, by
performing a series of radial velocity studies over a range of disc
precession phases -- is required to fully understand how large this
effect is on the mass estimate of the neutron star in Cyg~X-2.


\section*{Acknowledgments}

We acknowledge the use of the 1.5-m Tillinghast telescope at Fred L. Whipple Observatory, Mt. Hopkins, Arizona,
and we thank observers M.~L.~Calkins, P.~Berlind, T.~M.~Currie, E.~E.~Mamajek and W.~Peters.
We thank S.~Tokarz, N.~Caldwell and N.~Martimbeau for assistance with the sky line correction to the wavelength calibration.
This research made use of NASA's Astrophysics Data System, and the SIMBAD database, operated at CDS, Strasbourg, France.
This publication makes use of data products from the Two Micron All Sky Survey, which is a joint project of the
University of Massachusetts and the Infrared Processing and Analysis Center/California Institute of Technology,
funded by the National Aeronautics and Space Administration and the National Science Foundation.
We thank J. A. Orosz for providing us with the \elc\ code. We acknowledge the use of \molly\ and \doppler\ software
packages developed by T.~R.~Marsh, University
of Warwick.
We thank the anonymous referee for the helpful suggestions which improved our paper.
PE and PJC acknowledge support from Science Foundation Ireland.
MRG acknowledges partial support from NASA contract NAS8-03060 to the Chandra X-ray Center.

\appendix

\bsp

\label{lastpage}

\end{document}